# SERIMI – Resource Description Similarity, RDF Instance Matching and Interlinking

Samur Araujo[1], Jan Hidders[1], Daniel Schwabe[2], Arjen P. de Vries[1]

[1] Delft University of Technology, PO Box 5031, 2600 GA Delft, the Netherlands
{S.F.CardosodeAraujo, A.J.H.Hidders, A.P.deVries}@tudelft.nl

[2]Informatics Department, PUC-Rio Rua Marques de Sao Vicente, 225, Rio de Janeiro, Brazil
dschwabe@inf.puc-rio.br

**Abstract.** The interlinking of datasets published in the Linked Data Cloud is a challenging problem and a key factor for the success of the Semantic Web. Manual rule-based methods are the most effective solution for the problem, but they require skilled human data publishers going through a laborious, error prone and time-consuming process for manually describing rules mapping instances between two datasets. Thus, an automatic approach for solving this problem is more than welcome. In this paper, we propose a novel interlinking method, SERIMI, for solving this problem automatically. SERIMI matches instances between a source and a target datasets, without prior knowledge of the data, domain or schema of these datasets. Experiments conducted with benchmark collections demonstrate that our approach considerably outperforms state-of-the-art automatic approaches for solving the interlinking problem on the Linked Data Cloud.

## 1   Introduction

This technical report is an extended version of the paper that we submitted to ISWC2011 [32].

The interlinking of datasets published in the Linked Data Cloud (LDC) [27] is a challenging problem and a key factor for the success of the Semantic Web. Given the heterogeneity of the LDC, techniques aimed at supporting interlinking should ideally operate automatically and independently of a specific domain or schema.

In this context, ontology matching [10, 11, 15] (ontology alignment) and instance matching are the two most-studied sub-problems of interlinking. The former refers to the process of determining correspondences between ontological concepts. The latter often refers to the process of determining whether two descriptions refer to the same real-world entity in a given domain. Although these two problems are related, they are neither necessary nor sufficient to solve each other. For instance, we can have a perfect match between instances from two distinct datasets, for which there exists no ontology alignment. In this paper we focus on the problem of instance matching.

Our solution for the instance-matching problem is composed of two phases: the selection phase and the disambiguation phase. In the selection phase, for each instance **r** in a dataset A, we search for instances in a target dataset B that may refer to the same entity in the real world as **r,** by using a literal type of matching that has a high recall but a low precision. For example, this can be achieved by searching for instances in B that have a label similar to the label of the instance **r**. Since more than one instance in B may share the same label, the disambiguation phase is needed. In this phase, we attempt to filter among the instances found in B, those that actually refer to the same entity in the real world as **r**. Both phases assume that a higher similarity between two resources implies a higher probability that the two resources refer to the same entity or belong the same class.

The main question that arises is how to define the similarity between resources in each phase. The use of string distance measures (e.g. Jaro-Winkler [5], Levenshtein [6], Hamming [28]) can be used for determining similarity in the first phase of the problem (matching "labels"). In the second phase, the challenge is to define a function that uses an appropriate notion of similarity for the RDF data model. Many techniques exist in the literature for instance matching, but only a few of them (e.g. RiMOM [30] and ObjectCoref [29]) were applied and evaluated in the context of RDF instance matching. RiMOM applies a mix of matching strategies for solving the interlinking problem that, including the use of schema information to support the matching between instances. Mainly, it calculates the similarity based on the edit distance and cosine similarity, mixing information in the data and schema. ObjectCoref uses a semi-supervised machine learning technique that learns the discriminability of pairs of properties between the source and target instances. The notion of similarity between the source instances and target instances is based on the fact that two resources share a value in one of the properties in such a pair. Both systems are effective when the datasets being matched have a high degree of string similarity in their discriminative properties. Although they achieve a reasonable precision and acceptable recall on average, there are still a large number of cases in which these approaches do not work, as we will show in this paper.

Here we introduce a method called SERIMI. Our approach uses state of the art string matching algorithms for solving the selection phase, with an innovative function of similarity for approximating the notion of similarity during the disambiguation phase. This function is designed to operate even when there is no direct ontology alignment between the source and target datasets being interlinked. For example, SERIMI is able to interlink a dataset A that describes social aspects of countries with a dataset B that describes geographical aspects of countries. Our approach shows a considerable improvement in precision and recall when compared to state of the art tools for RDF instance matching and interlinking.

This paper is organized as follows. After this introduction, we discuss related work in Section 2. In Section 3, we introduce our approach for interlinking. In Section 4, we elaborate our method for detecting similarity between RDF resources. Section 5 presents a full description of our method for solving the interlinking problem. Section 6 and 7 present an experimental validation of the approach, with a comparative study against RiMON and ObjectCoref. Finally, Section 8 concludes this paper.

## 2   Related Work

Within data integration, a specific sub-problem is that of instance matching, which aims to bring together data from different and heterogeneous data sources and detect whether representations of entities contained represent the same entity in the real world. Instance matching [7] (also known as object matching, instance consolidation, duplicate identification, record linkage, entity resolution or reference reconciliation) is a crucial task in the process of data integration. This problem has been extensively studied in the field of databases in the last 20 years (recently surveyed in [1]).

The instance-matching problem is starting to attract attention from the Semantic Web community, since correctly establishing interlinks [2, 3] between RDF datasets is a critical success factor for the Semantic Web, especially for evolving the LDC. The RDF interlinking problem involves a particular setting of the instance-matching problem: the task is to match RDF resources, instead of database records, that represent the same entity in the real world. Although the RDF interlinking problem shares some characteristics with the record linkage problem in the database field, they differ in many aspects, caused by the flexibility of the RDF model.

Instance-matching techniques for database records assume that records are instantiations of a pre-defined database schema, which imposes the restriction that record properties (columns in a table) are fixed. This is not valid for RDF, since some ontologies (RDF schemas) impose no restriction on the use of properties, which implies that different approaches are necessary for the latter case.

Another aspect differentiating is that in database instance matching, the records to be matched should have the same table structure, and variation in the schema level is usually resolved before data integration starts. In RDF instance matching, two resources may be linked without necessarily having any of their properties in common. This means that, in RDF, the schema matching (e.g. ontology alignment) is not always necessary and sufficient for achieving RDF instance matching and interlinking.

Some techniques for instance matching are focused on matching names of the people, events or places [8, 9], and not the records themselves. Records and RDF resources are structured objects, which demands different matching strategies than those for matching simple strings.

The ontology-matching problem mentioned in the literature [13, 14, 15], refers to RDF data that represents a taxonomy or a vocabulary. In these cases the task is to match concepts described in two different ontologies. Although there are approaches that exploit the schema instance to find matches between taxonomies or vocabularies, the nature of these problems are different. In the interlinking over RDF data, the matches occur between instances that represent individual entities in the real world (e.g., John, Marie, Brazil, France, etc.), rather than between classes of entities (e.g. People, Country, etc).

The main issue in database and RDF instance-matching problems is the definition of a model that describes what is going to be matched. A classical work from the field of statistics has proposed a formal model for defining when two records match [7]. They formulate entity matching as a classification problem, where the goal is to classify whether entities match or not. In their definition they represent pairs of entities as pairs of structured objects that are represented as a vector of features. Two entities match if they have some degree of similarity in their vectors of features. Then, the challenge is to produce a function that computes the similarity between two vectors of features. In the RDF world there are two aspects that can be measured to produce a similarity measure between two RDF resources: the syntactical and semantic aspects. The syntactical aspect can be considered by applying string-similarity functions (Jaro-Wrinkler, Levenshtein) or set similarity (Cosine [24], Dice [20, 21], Jaccard [23]). These techniques work often well but fail when the syntactical similarity is not encountered in the data. For instance, we may have a resource A with a property *population* and another resource B with a property *number-of-inhabitants*, which semantically means the same but is syntactically completely different. In these cases, a semantic similarity function [16, 17, 18, 19] has to be applied in order to match these attributes, where a semantic similarity function is a function that uses external sources of knowledge (e.g. a dictionary, or an upper level ontology) during the process of matching.

In order to elaborate a methodology to detect the syntactical and semantical similarity in the context of instance matching in the Semantic Web, three main techniques are listed in the literature: manually constructed rules, supervised and unsupervised learning approaches

Techniques based on manually constructed rules [4] require a clear understanding of the schemas of the source and target datasets to be interlinked. The lack of knowledge about these schemas may lead the publisher to write excessively restrictive rules, therefore generating less interlinks than possible; or to write excessively inclusive rules leading to ambiguity among the selected resources, which decreases the precision of the interlinking. The supervised techniques demand the construction of a training set, which is a non-trivial task. It is almost impossible to conceive a generic training set to cover the whole heterogeneous universe of Linked Data. Unsupervised techniques mix string-matching algorithms with other techniques, such as machine learning or knowledge-based approaches [30].

In this paper we describe SERIMI, an unsupervised approach for solving the interlinking problem; and we compare it with RiMOM and ObjectCoref, both unsupervised techniques that try to solve the same problem.

## 3 Overview of RDF Instance Matching and Interlinking with SERIMI

A common scenario not addressed in the related work is the interlinking of sources with complementary information about the same entities. For instance, a resource can describe geographical aspects of the country Brazil, and another can describe social aspects of the country Brazil. By interlinking them, we can have an aggregated view about both aspects of the country Brazil. This disjoint representation of the same entity can also occur when we have data where its schema was automatically generated and the terms used are not human-readable. Since the previous scenarios occur quite frequently in the Semantic Web, we have designed an unsupervised technique, SERIMI that allows instance matching in the aforementioned situations.

In the context of Linked Data, the problem of instance matching can be defined as follows: given two distinct RDF datasets A and B, find pairs of resources, one from A and one from B, that refer to the same entity in a given domain. The process of finding those correspondences we called *instance matching*. The result of these mappings we will refer from now on as *interlinking*.

In SERIMI, the RDF interlinking involves several phases:

*Entity label property selection:* in order to select resources in the target dataset that can match a specific source resource, we first select the labels that represent these source resources.

*Pseudo-homonyms resources selection*: we can use these labels for searching for resources in the target dataset that share the same or similar labels. The set of resources that share a similar label, we will refer to *pseudo-homonym set*. An algorithm should be able to detect the properties in the source RDF dataset that play the role of these labels, without prior knowledge of the data or schema. We call these properties *entity label properties*, and their values *entity labels*. Once the entity labels properties are defined, we can use the *entity labels* of a source resource for searching for pseudo-homonym resources within the target dataset. Each pseudo-homonym resource has at least one value similar to the searched entity labels. This matching process is based on string matching algorithms. Although this process is necessary, it can lead to erroneous matching, since true positive matches may share exactly the same entity label than false positive matches. Therefore, we need a step for resolving this ambiguity in the pseudo-homonym set.

*Pseudo-homonyms resource disambiguation:* in some cases, a pseudo-homonym set may have instances of different classes or instances of the same classes that share the same label. We can illustrate these cases using the *Geonames Linked Data[1]*, where are resources labeled Brazil, which are instances of different classes, such as, city, street and hotel, including the country Brazil; as well as are resources labeled Cambridge that are instance of the same class (city) but refer to distinct locations. This fact leads to an ambiguity problem, in which we have to select among the pseudo-homonym resources, those that are more similar to the source resource.

To solve this ambiguity problem, we propose an innovative model called *Resource Description Similarity,* or *RDS*. The key idea is to complement the direct matches between the source and target datasets with the similarity relationships among the pseudo-homonym sets. RDS uses the intuition that if we select two or more resources that are similar in the source dataset, and for each of them there is a set of pseudo-homonyms resources in the target dataset, then the solution for each pseudo-homonym set are similar among themselves. It indicates that the solution to the problem is the set of resources that are more similar among pseudo-homonym sets. Fig.1 illustrates this intuition. Hence, the challenge is to define a model that approximates the notion of similarity in this fashion.

In the next sections we will detail the RDS model, to approximate the intuition above; and further we will prove empirically that the RDS model outperforms the state-of-art algorithms for solving the interlinking problem.

---

[1] http://www.geonames.org/ontology/documentation.html

|                          |                              |                          |
|--------------------------|------------------------------|--------------------------|
| Entity Label Brazil      | Entity Label Portugal        | Entity Label Spain       |
| Brazil as country        | Portugal as country          | Spain as country         |
| Brazil as river in Africa| Portugal as river in Africa  | Spain as city in Africa  |
| Brazil as river in Asia  | Portugal as city in America  | Spain as city in Europe  |
| Pseudo-homonym Set A     | Pseudo-homonym Set B         | Pseudo-homonym Set C     |

**Fig. 1** – A simple example of pseudo-homonym sets for three labels that represent countries.

## 4 Resource Description Similarity

The ambiguity problem can be isolated from the full interlinking process described before. It can be formally posed as:
Given:
1. a set of entity labels **EL** that each identify a certain entity in a certain class of interest **K** and together are representative for **K**,
2. and a target RDF dataset G,

Try to find for each label in **EL** the resources in G that represent the corresponding entity in **K**.

The crux of the ambiguity problem lays on determining the class of interest **K** (e.g. country in the example of Fig. 1). IN SERIMI, the determination of **K** narrows the size of each pseudo-homonym set to smaller set of resources, which by definition is the solution for the ambiguity problem.

### 4.1. Preliminary Definitions

In order to explain how the similarity function in our RDS model works, we need to give some preliminary definitions.

**Definition 1 (RDF Graph):** An RDF graph $G$ is s a set of triples, each of the form $(s, p, o)$ where $s \in U \cup B$, $p \in U$ and $o \in U \cup B \cup L$. Here $U$ is the set of *Uris* (concrete identifiers aka. *resources*), $L$ the set of *Literals* (basic strings) and $B$ the set of *Blank Nodes* (abstract identifiers). Within a triple $t = (s, p, o)$, $s$ is called the *subject*, $p$ the *predicate* and $o$ the *object* of $t$.

**Definition 2 (Datatype Property):** a predicate p is called a datatype property in RDF graph $G$ if for every triple (s, p, o) in $G$ it holds that $o \in L$.

**Definition 3 (Object Property):** a predicate p is called an object property in RDF graph $G$ if for every triple (s, p, o) in $G$ it holds that $o \in U$.

**Definition 4 (Inverse Functional Property):** a predicate p is said to be an inverse functional property, or IFP, in an RDF graph $G$ if for all triples (s, p, o) and (r, p, o) in $G$ it holds that $s=r$. The set of all IFP of $G$ is denoted as IFP($G$).

Notice that based on Definition 1, a resource $r \in U$ (a URI), does not necessarily contain enough information to be compared to another resource. In order to compute the similarity between resources in $G$, we need to select information about each resource $r$ that will be used in the comparison, which we call *the description* of $r$ in $G$. Therefore, we define a function *DF* that selects this information for a set of RDF resources from the graph $G$.

**Definition 5** (*Description Function*): the description function $DF: G \times 2^U \rightarrow G$ is defined such that $DF(G, W) = \{ (s, p, o) \mid (s, p, o) \in G \land s \in W \}$.

Informally, *DF*(*G*, *W*) retrieves the description of all resources in *W*. For instance, *DF*(G, {*r*}) returns all triples with subject *r*, in G. Notice that a different definition of the function *DF* may be applied as well. In this work we consider the function *DF* as defined in Definition 5.

We now formally define the pseudo-homonym set. For this purpose we first postulate a similarity relation over literals, denoted as *o ~ o'*. This similarity relation could be, for example, the *substring* relationship or the *Jaro-Winkler distance*.

**Definition 6** (*Pseudo-homonym Set of an Entity Label*): Given a graph *G*, and an entity label *e* in *EL*, we define the set of pseudo-homonyms of *e* as the *PH*(*e*) ={ *r* | ∃ *o* ∈ *EL*: *o ~ e* ∧ *(r, p, o)* ∈ *G* }.

Informally, the pseudo-homonym set of a label *e* is formed by all resources that are similar to *e* based on the value of a property *p* (in practice *p* is usually equals to predicates like: *rdf:label*, *foaf:name*, etc).

Lets illustrate the use of this function with an example. The pseudo-homonym set for the entity label "Brazil", i.e. *PH*("Brazil"), in DBPedia, would contain among others, the resources:

1. http://dbpedia.org/resource/Category:Brazil
2. http://dbpedia.org/resource/Brazil
3. http://dbpedia.org/resource/Colonial_Brazil
4. http://dbpedia.org/resource/Empire_of_Brazil
5. http://dbpedia.org/resource/Flag_of_Brazil
6. http://dbpedia.org/resource/Corcovado

In this example, we considered as *p*, the set of all properties available in the DBPedia dataset.

### 4.2. Overview of the RDS Function

Our solution to the ambiguity problem is summarized below:
1) Considering a given set **EL** of entity labels, find the pseudo-homonym set PH(e) for each entity label e in **EL**;
2) given the sets of pseudo homonyms PH($e_1$), ... , PH($e_n$), for each resource in each pseudo-homonym set, compute a δ similarity measure for it as defined by the RDS-FUNCTION below;
3) based on these δ, classify if a resource belong to the class of interest **K**.

The pseudo-code of this procedure is shown in the Fig. 2a, where the set **EL** is denoted by $\{e_1, ..., e_n\}$.

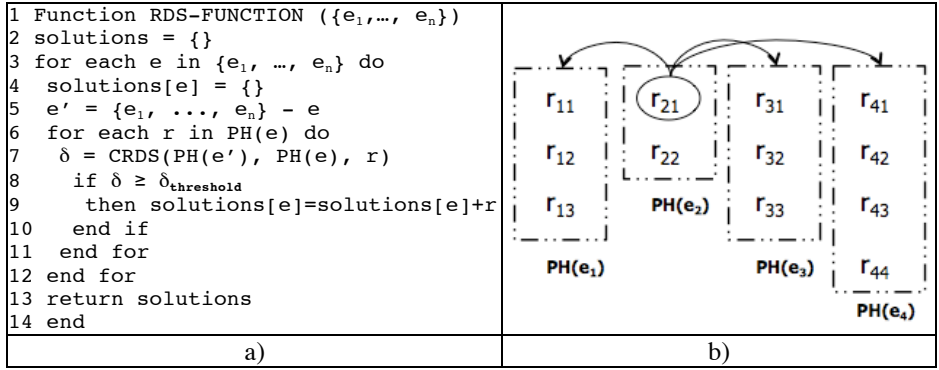

```
1 Function RDS-FUNCTION ({e₁,…, eₙ})
2 solutions = {}
3 for each e in {e₁, …, eₙ} do
4   solutions[e] = {}
5   e' = {e₁, ..., eₙ} - e
6   for each r in PH(e) do
7     δ = CRDS(PH(e'), PH(e), r)
8     if δ ≥ δ_threshold
9       then solutions[e]=solutions[e]+r
10    end if
11   end for
12 end for
13 return solutions
14 end
```

a)                                                                b)

**Fig. 2** – a) Pseudo code for the RDF-FUNCTION, b) *RDS-FUNCTION* compare a resource to all others pseudo-homonyms sets. For the resource $r_{21}$ (and $r_{22}$), we compare it with $PH(e_1)$, $PH(e_3)$ and $PH(e_4)$

Fig.2b depicts the loop described in the line 6 of Fig. 2 (for the resource $r_{21}$). The definition of CRDS will be given further on. The idea is to classify a target resource as belonging to the class of interest **K** by comparing it to all the other pseudo-homonym sets. This comparison generates a score δ. If $δ ≥ δ_{threshold}$, we conclude that r is of the class **K** and therefore the match of e. We will explain how we obtained this threshold further on in this paper, which is estimated automatically.

### 4.3. The RDF Resource Similarity Function in the RDS Context

In order to build the CRDS function we need to define how we compute the similarity between sets of resources.

**Definition 7** (*Items of Measurement*): given a graph *G* and a set of resources *X* in *G*, we defined the sets:
- $P(X) = \{p \mid (s, p, o) \in DF(G, X) \land s \in X\}$;
- $D(X) = \{o \mid (s, p, o) \in DF(G, X) \land s \in X \land o \in L\}$;
- $O(X) = \{o \mid (s, p, o) \in DF(G, X) \land s \in X \land o \in U\}$ and;
- $T(X) = \{(p,o) \mid (s, p, o) \in DF(G, X) \land s \in X\}$.

In the calculation of $D(X)$, $O(X)$ and $T(X)$, we eliminate triples that have predicates that occur, for the same subject, with a cardinality $C > \max(\eta, 5)$, where η is equal to the average of the cardinality C among all predicates from all triples encountered in *X*. This removes noisy triples from the data.

### 4.3.1. Items of Measurement

The choice of the items of measurement is strongly related to how we conceive similarity among RDF resources. The highest degree of similarity that two resources r and s can have is when for all triples (r, p, o) and (s, p, o); (r, p, o) = (s, p, o). In this case, r is equal to s. As a general rule when a triple has the subject r in common with the subject s of another triple, then r and s can be considered the same. However, we are not interested in detecting that two resources are the same, but in detecting the degree of similarity that two resources may have.

Note that each item of measurement may define distinct classes of interest. Some authors refer to this property as *salience* [31] of an item of measurement. In our approach we equally weighty the items of measurement, since we empirically observed that this configuration performs better. However, there are reasons to motive the use of different weight for different items of measurement, since different weight may influence the definition of the class of interest.

As we will produce an average measure of similarity the salience of each item of measurement will be equalized in the average. The similarity between two sets of resources *A* and *B* is given by:

$$RDS(A,B) = SetSim(P(A), P(B)) + SetSim(D(A), D(B)) \\ SetSim(O(A), O(B)) + SetSim(T(A), T(B))$$ (1)

### 4.3.2. Defining a Function of Similarity

Mainly, we want the SetSim to reflect the intuition that two sets A and B that have *n* features in common should be more similar than two sets with *n-1* features in common, no matter the number of distinct features both cases may have. SetSim[2] was derived from the Tversky's contrast model [12] and is given by equation 2:

$$SetSim(A,B) = |A \cap B| - \frac{|A - B|}{|A \cup B|}$$ (2)

Although there are several set-based similarity indexes (Jaccard, Dice, Cosine) available in the literature, there are a couple of reasons for designing a specific index adjusted to the SER approach. We will present our motivations in the next paragraphs and empirically prove that the SetSim index outperforms the existing ones further on in this paper.

---

[2] In our experiments SetSim beats the common Jaccard index by a small but consistent margin.

We decided to derive the SetSim from the *Tversky's contrast model*. In the contrast model the similarity between two entities, A and B, is expressed as a linear combination of the measures of theirs common and distinctive attributes, as shown in equation below.

$$Tversky(A,B) = \lambda f(A \cap B) - \alpha f(A - B) - \beta f(B - A) \qquad (3)$$

where, $\alpha$, $\beta$, and $\lambda \geq 0$.

The contrast model is composed by three disjoint set functions. The scale function $f(A \cap B)$ represents the set of common attributes between A and B. The function $f(A - B)$ and $f(B - A)$ represents the set of distinct attributes between A and B, and B and A, respectively. The constants $\alpha$, $\beta$, and $\lambda$ allows to weight independently the communalities and differences in the measure of similarity. In our case, the scale function $f$ in the contrast model denotes a function that takes a set x of attributes and returns its cardinality |x|.

The advantage of using the contrast model instead of the others model of similarity is because it captures the similarity of two resources, or set of them, based on theirs communality and distinctness. The use of distinctiveness is relevant in our case because communalities between resources are not enough to distinguish them. For example, suppose you want to find resources that represent the labels "Amazon", "Minas Gerais" in a certain dataset $G$. Since we assume that the given set of labels is representative for the class $K$, it is reasonable to assume that the class of interest $K$, in this case, is the class of *Brazilian states* - without consider any other information about the class $K$, Brazilian States are the most *salient* entities to represent these labels, from the cognitive point of view. Suppose also that in the target dataset $G$ there are two resources that matches with "Amazon": the state of Amazon in Brazil and the Amazon River in Brazil; and there is a resource the state of Minas Gerais that matches with the label "Minas Gerais". Suppose also that these resources have a unique property in common that indicates that they are located in Brazil. If we judge theirs level of similarity just by taking into account the fact they are located in Brazil, we would arrive to the conclusion that all of them are similar in the same degree. Consequently, it would lead us to select the Amazon River and the State of Amazon as representative resources for the string "Amazon". However, clearly the Amazon River is not an instance of the class of Brazilian states. The distinction between Amazon River from a Brazilian state can be detected whether we consider the differences that these entities have. Therefore, this example shows how the contrast model can capture the distinctiveness factor, which produces a more accurate measure of similarity.

### 4.3.3. Tuning parameter in the contrast model

By tuning $\alpha$ and $\beta$ parameters in the contrast model, we can attribute more or less weight to the difference factors in the measurement. As we evidenced empirically, the general rule is to have more distinct value than common values among the resources compared, in the context of SERIMI. For this reason, we decided to attribute less weight to the difference factors in contrast model. Therefore, the communality will prevail in the measure of similarity. Notice that Jaccard and Dice's give the same degree of importance for the distinctiveness and communality factors in the measure, therefore producing a measure biased towards the difference, due to the characteristic of the data in our case.

In the SERIMI approach, for all resources in a pseudo-homonym set R, such as R $\in$ S, we compare them to all pseudo-homonym set S'$\in$ S–R. In this way, the parameter A always refers to a singleton set containing a resource r from R and B refers to a pseudo-homonym set S'. For this reason, the difference factor between B–A is not relevant in the final measure, since given two resources s and r in R if |T(s) $\cap$ T(S')| = |T(r) $\cap$ T(S')| then |T(s) - T(S')| = |T(r) - T(S')|. The factor B-A does not add any relevant information to the final measure, and for this reason we defined it as:

$$\alpha = 0$$

Although the factor B–A is not relevant, the factor A–B adds relevant information to the measure. However, we want that this factor to have a lower weight in the measure than the commonality factor A$\cap$B, therefore this latter factor prevail over the former. Consequently, the parameters $\beta$ is defined as:

$$\beta = |A \cup B|^{-1}$$

These parameters produces an accurate measure of similarity, not biased by the cardinality of B–A or A–B. Therefore, we can derive the SetSim index from the Tversky' contrast model, where f(x) is the cardinality function |x|, and $\lambda = 1$, $\alpha = 0$ and $\beta = (|A \cup B|)^{-1}$

### 4.3.4. Resource Description Similarity Function

Equation 4 uses the RDS function to compute a score of similarity for each resource in the pseudo-homonyms sets. It compares a resource *r* to all other pseudo-homonym sets in which *r* does not belong to. Fig. 2b depicts this process. In equation 3, we divide the RDS({r}, S') by the cardinality of S', such that a set S' with high cardinality will affect less the final result of the measure.

$$URDS(r,S) = \sum_{S' \in \{T | T \in S \wedge r \notin T\}} \frac{RDS(\{r\},S')}{|S'|} \tag{4}$$

We normalize the results of equation 4 by the maximum measure among all resources in R, resulting in the similarity between the resources *r* and a set *S* of pseudo-homonym sets given by equation 5:

$$CRDS(r,R,S) = \frac{URDS(r,S)}{\max_{r' \in R} URDS(r',S)}, \quad r \in R, \ R \in S \tag{5}$$

Using equation 5, a resource is considered a solution to the RDS problem iff CRDS(r, R, S) is higher than a defined threshold δ**threshold** or its rank is within the Top-k. An appropriate value for $\delta_{threshold}$ is the maximum of the mean and median, over all δ of the resources in R, which we will refer as $\delta_m$. Although it is possible to define this value automatically by applying other score distribution functions [26], it demands an adjustment of these functions to our model, which is a research area by itself. For purpose of comparison, we will use in our evaluation: $\delta_{threshold} = \delta_m$, $\delta_{threshold} =$ 1.0, $\delta_{threshold} = 0.95$, $\delta_{threshold} = 0.90$ and $\delta_{threshold} = 0.85$. Also, we evaluated the Top-k approach, in which we selected the Top-k resources in R, ordered decreasingly by δ. We evaluated the Top-1, Top-2, Top-5 and Top-10 thresholds.

## 5 The SERIMI Algorithm

In this section we detail how we solve the interlinking problem using the SERIMI's approach. Fig. 3 summarizes our process.

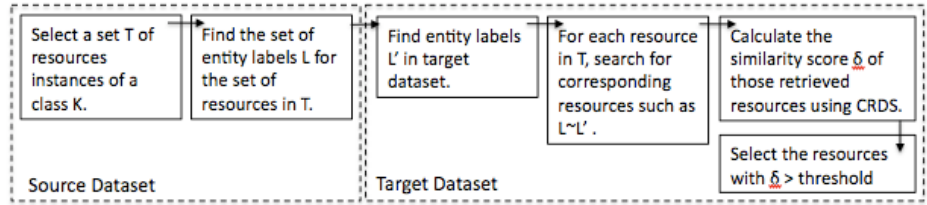

**Fig.3** – SERIMI's process to solve the interlinking problem.

*Selecting resources to interlink*: SERIMI operates over instances of a specific class of interest **K** in the source dataset. Once the class **K** is defined, it starts the process of interlinking described below. For the sake of this paper we assume that the class **K** is manually defined. Notice that in order to interlinking the whole source dataset, a trivial algorithm can be used for selecting the class **K** automatically. For example, we can obtain classes in RDF graph by selecting objects of triples with property *rdf:type*. Although not optimal, the use of this simple heuristic may work well for the Linked Open Data.

*Selecting entity label property*: the second step of the process is to select the entity label properties in the source dataset. We consider as entity label properties, all RDF predicates that have a literal value with less than 200 characters. Since predicates with higher entropy are more discriminative than predicate with lower entropy, we used that information to select the predicates with entropy $\Omega \geq \Omega_{threshold}$, where $\Omega_{threshold}$ is obtained by averaging the entropy of all predicates of the resources that we want to interlink. The entropy *H(P)* of a predicate *P* with possible literal values {$o_1$, .., $o_n$}, can be given by equation 6, where *p* is assumed to be the probability mass function of *P*. In the source dataset we execute it over the subset of instances selected for interlinking.

$$H(P) = -\sum_{i=1}^{n} p(o_i) \log_2 p(o_i) \qquad (6)$$

*Selecting corresponding resources*: Once we have determined the entity label properties, we can build the pseudo-homonym sets. For each instance to be interlinked, we select the entity label of the entity label property with the highest entropy. We use this entity label for searching for resources in the target dataset. We apply the same step described in the previous paragraph over the target dataset, which gives us a set of predicates. Then we search for the entity label only on triples that contain such selected predicates. Afterwards, we select the subject resource(s) with a maximum Jaro-Winkler with respect to the searched entity label. If the maximum score is below 70% we discard it. By selecting only those resources with maximum relative similarity measure, we reduce the number of resources in the pseudo-homonym set, thereby improving the chance of true positive matches. If no resource is retrieved, then we select the next entity label property with the highest entropy and repeat the same procedure.

*Disambiguating candidate resources*: once the pseudo-homonym sets are built using the procedure above, we can execute the RDS-FUNCTION. Given S as a set of all pseudo-homonyms and R ∈ S, for each resource r in R, we generate a score $\delta$ = CRDS(r, R, S). As solution for a pseudo-homonym set R, we select all resources with a score $\delta \geq \delta_{threshold}$.

### 5.1. Eliminating Outliers

Naturally, some of the CRDS predictions are not sufficiently reliable, since CRDS(r, R, S) can normalize even small values to 1. To optimize our result, we eliminate outliers that are below a specific threshold ϕ, before the normalization. A reasonable approximation of ϕ is given by the difference between the mean of Δ and its standard deviation σ; where Δ is the set of values attributed (by the equation 4) to the resources in *all* pseudo-homonym sets. We only consider this heuristic for Δ with standard deviation σ > 0.13, since we observed that cases with small σ do not have a bell-curve.

### 5.2. Making SERIMI Scalable

When the source dataset is large, the number of pseudo-homonyms sets to consider increases, affecting the computation time of CRDS. Therefore, we execute the process described so far sequentially over chunks of instances of size $\mu$, where $\mu \geq 2$. In order to determine an appropriate value for $\mu$, we evaluate different values. We tested the set of values {2, 5, 10 20, 50, 100}. The result of the performance of the method for each value is shown in the Section 7, Table 2.

Another advantage of the use of chunks is that for very loop of SERIMI, we can select the predicted resource with the highest δ and we add it as a singleton set to the set S of pseudo-homonym set in the next interaction. This extra set acts as a pivot in the function CRDS(r, R, S), increasing the evidence of the instances that belong to the class of interest **K**. We execute this procedure cumulatively, for every interaction and we add a total of $\mu$ (the size of the chunk) singleton sets to the set S. In this way, we give a reasonable amount of evidence to the CRDS, without delaying its performance too much, and thereby we improve the accuracy of our approach.

## 6 Experiment Setup

In this section we describe the employed collections, evaluation metrics and baselines applied in our evaluations.

### 6.1. Collections

We used the collection proposed in the instance-matching track of the Ontology Alignment Evaluation Initiative (OAEI 2010) [25]. We focused our evaluation in the life science (LS) collection (which includes DBPedia[3], Sider[4], Drugbank[5], LinkedCT[6], Dailymed[7], TCM[8], and Diseasome[9]) and in the Person-Restaurant (PR) collection proposed by this initiative.

The life science collection includes 7 datasets, which sum up hundreds of millions of triples. They contain several cases of ambiguity, which makes the interlinking problem hard to solve over them. We evaluated the same pairs of datasets that were evaluated by other systems and reported upon in the initiative. The Person-Restaurant collection contains 3 pairs of datasets. Two of these pairs describe people and the other pair describes restaurants. These datasets contain instances from one class, and together sum up a few thousands of resources. Table 1 shows the number of reference alignments for the pairs of datasets evaluated.

We have loaded all these datasets into an open-source instance of Virtuoso Universal server[10], where around 2GB of data were loaded. An exception was the DBPedia dataset, which we accessed online via its Sparql endpoint. The Virtuoso server was installed in a Mac OS X – version 10.5.8, with 2.4 GHz Intel Core 2 Duo processor and with 4 GB 1067 MHz DDR3 of memory. We ran the script that implements the SERIMI approach directly over the local SPARQL endpoints and DBPedia online endpoint.

---

[3] http://dbpedia.org/About
[4] http://www4.wiwiss.fu-berlin.de/sider/
[5] http://www4.wiwiss.fu-berlin.de/drugbank/
[6] http://data.linkedct.org/
[7] http://www4.wiwiss.fu-berlin.de/dailymed/
[8] http://code.google.com/p/junsbriefcase/wiki/RDFTCMData
[9] http://www4.wiwiss.fu-berlin.de/diseasome/
[10] http://virtuoso.openlinksw.com/dataspace/dav/wiki/Main/

**Table 1.** Number of reference alignment for the pair of datasets evaluated in the LS collection.

| Dataset Pair | N° of Ref. Alignment | Dataset Pair | N° of Ref. Alignment |
|---|---|---|---|
| Sider-DBPedia | 1509 | Dailymed-DBpedia | 2549 |
| Sider-Dailymed | 1634 | Dailymed-LinkedCT | 27729 |
| Sider-Diseasome | 173 | Dailymed-TCM | 33 |
| Sider-DrugBank | 1140 | Dailymed-Sider | 1592 |
| Sider-TCM | 171 | Drugbank-Sider | 284 |
| Diseasome-Sider | 238 | Person11-Person12 | 500 |
| Restaurant1-Restaurant2 | 112 | Person21-Person22 | 400 |

### 6.2. Evaluation metrics and baseline

In order to evaluate the effectiveness of the proposed interlinking method, we used the precision, recall and F1 metrics. These formulas are defined in the equation 7, 8 and 9, respectively.

$$recall = \frac{|true\_positive|}{|true\_positive| \cup |false\_negative|} \quad (7)$$

$$precision = \frac{|true\_positive|}{|true\_positive| \cup |false\_positive|} \quad (8)$$

$$F1 = \frac{2 * precison * recall}{precsion + recall} \quad (9)$$

We considered as true positives the reference alignment. False positives are the alignments found by SERIMI that do not exist in the reference alignment, the false negatives those not found by SERIMI.

We used two baselines in our experiments: RiMOM and ObjectCoref. These two systems are representative of the two main types of solution for the interlinking task and, more importantly, they have used the same set of datasets and reference alignment as our method, allowing a fair and direct comparison. In addition we investigated how SERIMI performs using different threshold for parameters δ and $\mu$. Consequently, we obtained a full picture of how each feature of our approach affects the overall performance of SERIMI.

# 7 Results

In this section we show the results of our experiments. Table 2 shows SERIMI's performance when we vary the parameter $\mu$. As we can see, the standard deviation of the precision and recall are close to zero, meaning that the different configurations, that is, the size of the chunk of source resources, does not affect much the SERIMI's overall performance. In the remainder, we decided to use $\mu = 20$, for a good trade-off between effectiveness and efficiency.

**Table 2.** SERIMI's precision (P) and recall (R) varying the parameter $\mu$.

| Dataset Pair | Sider-Dailymed | | Drugbank-Sider | | Person11-Person12 | |
| --- | --- | --- | --- | --- | --- | --- |
| Parameter $\mu$ | P | R | P | R | P | R |
| SERIMI, $\mu = 2$ | 0.805 | 0.539 | 0.330 | 0.915 | 0.991 | 0.986 |
| SERIMI, $\mu = 5$ | 0.799 | 0.558 | 0.329 | 0.912 | 0.986 | 0.986 |
| SERIMI, $\mu = 10$ | 0.799 | 0.558 | 0.327 | 0.905 | 0.992 | 0.992 |
| SERIMI, $\mu = 20$ | 0.780 | 0.576 | 0.329 | 0.915 | **1.000** | **1.000** |
| SERIMI, $\mu = 50$ | 0.825 | **0.595** | **0.331** | 0.922 | **1.000** | **1.000** |
| SERIMI, $\mu = 100$ | **0.834** | 0.590 | 0.330 | **0.922** | **1.000** | **1.000** |
| Standard Deviation | 0.020 | 0.021 | 0.001 | 0.006 | 0.006 | 0.006 |

Table 3 shows SERIMI's performance when we vary the threshold $\delta_{threshold}$. We observed that the standard deviation of the precision and recall is close to zero in cases (e.g. Drugbank-Sider) when we have too many pseudo-homonym sets with cardinality equal to 1. It is relatively high otherwise (e.g. Sider-Dailymed). This shows that different configurations, that is, the restriction on degree of similarity of the resources to be considered a solution, may affect the SERIMI's overall performance. Although not optimal, the use of the $\delta_m$ (as introduced in end of Section 4.3) as threshold performs relatively well in all cases. Therefore, for all other experiments shown in this paper, we used $\delta_{threshold} = \delta_m$.

**Table 3.** SERIMI's precision and recall varying $\delta_{threshold}$.

| Dataset Pair | Sider-Dailymed | | Sider-Drugbank | | Drugbank-Sider | | Person11-Person12 | |
| --- | --- | --- | --- | --- | --- | --- | --- | --- |
| $\delta_{threshold}$ | P | R | P | R | P | R | P | R |
| SERIMI, $\delta >= \delta_m$ | **0.780** | 0.576 | 0.973 | 0.969 | 0.328 | 0.915 | **1.000** | **1.000** |
| SERIMI, $\delta = 1.0$ | 0.706 | 0.368 | 0.762 | 0.364 | 0.328 | 0.910 | **1.000** | **1.000** |
| SERIMI, $\delta >= 0.95$ | 0.743 | 0.516 | 0.792 | 0.552 | 0.328 | 0.915 | 0.994 | 0.996 |
| SERIMI, $\delta >= 0.90$ | 0.771 | 0.673 | 0.808 | 0.687 | 0.328 | 0.915 | 0.972 | 0.996 |
| SERIMI, $\delta >= 0.85$ | 0.778 | 0.761 | 0.816 | 0.761 | 0.328 | 0.915 | 0.968 | 0.996 |
| Standard Deviation | 0.033 | 0.174 | 0.024 | 0.174 | 0.000 | 0.000 | 0.092 | 0.005 |
| Top-k | P | R | P | R | P | R | P | R |
| SERIMI, Top-1 | 0.545 | 0.280 | **0.978** | 0.973 | **0.330** | **0.919** | 0.986 | 0.984 |
| SERIMI, Top-2 | 0.581 | 0.430 | 0.966 | 0.979 | **0.330** | **0.919** | 0.978 | 0.992 |
| SERIMI, Top-5 | 0.633 | 0.683 | 0.966 | **0.980** | **0.330** | **0.919** | 0.969 | 0.996 |
| SERIMI, Top-10 | 0.661 | **0.848** | 0.966 | **0.980** | **0.330** | **0.919** | 0.967 | 0.996 |
| Standard Deviation | 0.052 | 0.254 | 0.006 | 0.003 | 0.000 | 0.000 | 0.009 | 0.006 |

In Table 4 we show the performance of SERIMI in solving interlinking for all pairs of datasets discussed in Section 6, in comparison with our two baselines: RiMOM and ObjectsCoref.

**Table 4.** It shows SERIMI, RiMOM and ObjectCoref the precision and recall for all dataset pairs. ObjectCoref's results are not available for all pairs of datasets.

| Dataset Pair / Approaches | Sider-DBPedia | | Sider-Dailymed | | Sider-Diseasome | | Sider-DrugBank | | Sider-TCM | |
|---|---|---|---|---|---|---|---|---|---|---|
| | P | R | P | R | P | R | P | R | P | R |
| SERIMI | 0.479 | **0.617** | **0.780** | 0.567 | **0.922** | 0.831 | **0.973** | **0.969** | **0.970** | **0.976** |
| RiMOM | **0.717** | 0.482 | 0.567 | **0.706** | 0.315 | **0.837** | 0.961 | 0.342 | 0.778 | 0.812 |
| ObjectCoref | - | - | - | - | - | - | - | - | - | - |

| Dataset Pair / Approaches | Dailymed-DBpedia | | Dailymed-LinkedCT | | Dailymed-TCM | | Dailymed-Sider | | Drugbank-Sider | |
|---|---|---|---|---|---|---|---|---|---|---|
| | P | R | P | R | P | R | P | R | P | R |
| SERIMI | **0.611** | **0.330** | **0.234** | 0.051 | **0.233** | **0.911** | 0.541 | 0.868 | **0.328** | 0.915 |
| RiMOM | 0.246 | 0.293 | 0.070 | **0.235** | 0.159 | 0.535 | **0.567** | 0.706 | - | - |
| ObjectCoref | - | - | - | - | - | - | 0.548 | **0.999** | 0.302 | **0.996** |

| Dataset Pair / Approaches | Diseasome-Sider | | Person11-Person12 | | Person21-Person22 | | Restaurant1-Restaurant2 | |
|---|---|---|---|---|---|---|---|---|
| | P | R | P | R | P | R | P | R |
| SERIMI | 0.833 | **0.901** | **1.000** | **1.000** | 0.557 | 0.385 | 0.768 | 0.768 |
| RiMOM | - | - | 1.000 | 1.000 | 0.952 | **0.990** | 0.860 | 0.768 |
| ObjectCoref | **0.837** | 0.668 | 1.000 | 0.998 | **0.989** | 0.887 | **0.989** | **0.804** |

As we can see in Table 4, SERIMI outperforms the baselines in 70% of the cases. The poor performance of SERIMI on Dailymed-DBPedia, Dailymed-Linkedct, Dailymed-TCM, Drugbank-Sider and Sider-DBPedia may be attributed to inaccuracy in the reference alignment. Although this issue is more evident in those cases, as also reported by RiMOM and ObjectCoref authors, nearly in all reference alignment we encountered missing alignments, inconsistencies, and even evidences of incorrect alignment. Below we show some examples of missing alignments in the reference alignment.

- *Sider-Dailymed:*

    **Source resource:**
    http://www4.wiwiss.fu-berlin.de/sider/resource/drugs/151165
    **Reference alignment:**
    http://www4.wiwiss.fu-berlin.de/dailymed/resource/drugs/1050

    **Missing alignment:**
       http://www4.wiwiss.fu-berlin.de/dailymed/resource/drugs/1618
- *Dailymed-DBPedia:*

   **Source resource:**
   http://www4.wiwiss.fu-berlin.de/dailymed/resource/organization/Allergan,_Inc.
   http://www4.wiwiss.fu-berlin.de/dailymed/resource/organization/AstraZeneca
   http://www4.wiwiss.fu-berlin.de/dailymed/resource/organization/B._Braun
   **Missing alignment:**
   http://dbpedia.org/resource/Allergan
   http://dbpedia.org/resource/AstraZeneca
   http://dbpedia.org/resource/B._Braun_Melsungen

This reflected negatively in the precision and recall of all approaches. Apart of the results shown in Table 4, we manually fixed the Dailymed-TCM reference alignment and recalculated the precision and recall, obtaining 100% precision and 98% recall for this case. This new alignment and SERIMI implementation are available for download[11].

The poor performance of SERIMI in the Restaurant1-Reataurant2 is mainly due to missing alignment in the reference set. The poor performance in the Person21-Person22 pair is due to the nature of the data. These datasets where built by adding spelling mistakes to the properties and literals values of their original datasets. Also only instances of class Person were retrieved into the pseudo-homonym sets during the interlinking process. Although these factors did not influence negatively the selection phase of the interlinking process, both together did not provide enough information for the CRDS function to generate distinguishable scores, consequently reflecting in its inaccurate predictions, in this case. Since ObjectCoref and RiMOM were able to exploit better the similarity in misspelled data and provide accurate predictions, it shows that this is a good area to explore as future work.

Table 5 shows SERIMI's performance when we vary the type of similarity. We evaluated SERIMI with SetSim and Jaccard index, and we averaged the F1 of all pairs of datasets shown in Table 1. As we can see, the SetSim index indeed performs the highest F1 metric. Although not visible in these figures, we observed that the difference between the indexes is more evident when pseudo-homonym sets have high cardinality. For instance, in the pair Sider-Dailymed, SetSim index outperformed Jaccard index by 50% under the F1 metric. It confirms our intuition that the difference should not prevail in the measurement of similarity in the context of RDS. The Dice's index produces exactly the same results than Jaccard index for the case that we evaluated, also motivated by [22].

**Table 5.** It shows the SERIMI's averaged F1 over the 13 pairs of datasets when we vary the type of similarity.

| Metric/Index | F1 |
|---|---|
| SetSim Index | **0.650** |

---

[11] https://github.com/samuraraujo/SERIMI-RDF-Interlinking

| | |
|---|---|
| Jaccard Index | 0.622 |

## 8   Conclusion and Future Work

RDF instance matching in the context of interlinking RDF datasets published in the Linked Data Cloud is the task of determining if two resources are referred to the same entity in the real world. This is a challenging task in high demand by data publishers that wish to interlink their datasets in the cloud.

In this work, we propose a novel approach, called SERIMI, for solving the RDF instance-matching problem automatically. SERIMI matches instances between a source and target datasets, without prior knowledge of the data, domain or schema of these datasets. It does so by approximating the notion of similarity by pairing instances based on entity labels as well as structural (ontological) context. As part of the SERIMI approach, we proposed the CRDS function to approximate that judgment of similarity.

We used two collections proposed by the OAEI 2010 initiative to evaluate SERIMI. On average, SERIMI outperforms two representative systems, RiMOM and ObjectCoref, which tried to solve the same problem using the same collections and reference alignment, in 70% of the cases.

As future work, we intend to investigate how our model can be adjusted to consider partial string matching in the similarity function that we proposed, and to accommodate different score distribution metrics as the threshold for the parameter $\delta_{threshold}$. Also, we intend to evaluate this approach in different collections that may provide a more accurate reference alignment than the ones that we used in this work.